\documentclass{aa}
\usepackage{graphicx}

\usepackage{natbib} 
\bibpunct{(}{)}{;}{a}{}{,} 

\usepackage{hyperref}

\usepackage[english]{babel} 

\begin{document}

   \title{Analytic modeling of recurrent Forbush decreases caused by corotating interaction regions}


   \author{B. Vr\v{s}nak\inst{1}
              \and
           M. Dumbovi\'c\inst{1}
              \and
           B. Heber\inst{2}
             \and
           A. Kirin \inst{3}
          }

   \offprints{B. Vr\v{s}nak, \email{bvrsnak@geof.hr}}

   \institute{
   Hvar Observatory, Faculty of Geodesy, University of Zagreb, Ka\v{c}i\'{c}eva 26, HR--10000 Zagreb, Croatia
                \and
   Institut f{\" u}r Experimentelle und Angewandte Physik, Christian-Albrechts-Universit\"at zu Kiel, Christian-Albrechts-Platz 4, 24118 Kiel, Germany
      \and
   Karlovac University of Applied Sciences, Trg J.J. Strossmayera 9, 47000 Karlovac, Croatia
             }

   \date{Received...................; Accepted ....................}


  \abstract
   {On scales of days, the galactic cosmic ray (GCR) flux is affected by coronal mass ejections and corotating interaction regions (CIRs), causing so-called Forbush decreases and recurrent Forbush decreases (RFDs), respectively.}
   { We explain the properties and behavior of RFDs recorded at about 1\,au that are caused by CIRs generated by solar wind high-speed streams (HSSs) that emanate from coronal holes.  }
   {We employed a convection-diffusion GCR propagation model based on the Fokker-Planck equation and applied it to solar wind and interplanetary magnetic field properties at 1\,au.  }
   {Our analysis shows that the only two effects that are relevant for a plausible overall explanation of the observations are the enhanced convection effect caused by the increased velocity of the HSS and the reduced diffusion effect caused by the enhanced magnetic field and its fluctuations within the CIR and HSS structure. These two effects that we considered in the model are sufficient to explain not only the main signatures of RFDs, but also the sometimes observed ``over-recovery'' and secondary dips in RFD profiles. The explanation in terms of the convection-diffusion GCR propagation hypothesis is tested by applying our model to the observations of a long-lived CIR that recurred over 27 rotations in 2007-2008.}
   {Our analysis demonstrates a very good match of the model results and observations.}


\authorrunning{B. Vr\v snak et al.}
\titlerunning{CIR-associated recurrent Forbush decreases}

   \maketitle
%

\section{Introduction}

Recurrent high speed streams (HSSs) in the solar wind that originate from coronal holes cause various important effects in the heliosphere
    \citep[e.g.,][and references therein]{nolte76,schwenn06,tsurutani06,vrsnak07,rotter12,hofmeister18};
for a review, see \citet{cranmer09}.
The interaction of HSSs with the preceding slow solar wind creates the so-called corotating interaction regions \citep[CIRs; e.g.,][and references therein]{gosling99,schwenn06,tsurutani06}; for a review, see \citet{richardson18}.
These HSS and CIR structures cause recurrent moderate geomagnetic storms and substorms, where the
Disturbance Storm-Time  index (Dst) rarely goes below $-100$\,nT \citep[e.g.,][]{tsurutani06,vrsnak07,zhang07,Verbanac11a,Verbanac11b,Verbanac13,vrs17},
as well as temporal depressions in the galactic cosmic ray (GCR) count rate,
   which are generally known as recurrent Forbush decreases \citep[RFDs; for a review, see, e.g.,][]{iucci79,kunow95,richardson04,richardson18}.
   Generally, Forbush decreases \citep[FDs;][]{forbush37} are temporal, well-defined decreases in the GCR count rate that are caused by interplanetary coronal mass ejections (ICMEs) and CIR structures \citep[e.g.,][and references therein]{lockwood71,cane00,richardson04,dumbovic11,dumbovic12,richardson18};
for reviews, see \citet{cane00} and \citet{richardson04}.
   We note here that some authors exclusively use the term Forbush decrease for GCR depletions caused by ICMEs, and those that are caused by CIRs, they call ``recurrent GCR depressions''.
RFDs are usually relatively weak, and their amplitudes are considerably lower than those of ``standard'' FDs caused by ICMEs
   \citep[e.g.,][and references therein]{richardson04,calogovic09,dumbovic11,dumbovic12,badruddin16,melkumyan19}.

In recent years, a number of studies related to RFDs were published. For example, \citet{kumar14}, \citet{badruddin16}, \citet{ghanbari19}, and \citet{guo21} studied the behavior of RFDs by applying the superposed epoch analysis approach. We also refer to the observation-based studies by  \citet{modzelewska20} and \citet{shen20}, as well as to the numerical studies by \citet{alania11}, \citet{guo16}, \citet{kopp17}, and \citet{luo20}.

The interaction of the high-speed stream with the preceding slow solar wind creates the CIR structure by compressing the plasma in the region in which they interact \citep[][]{gosling99,jian06,richardson18}. The compression region includes parts of the slow solar wind and parts of the HSS, divided by the so-called stream interface. The compressed interaction region is characterized by a continuously increasing flow speed, consistent with the continuity equation. The compression region is followed by the rarefaction region, in which the flow velocity gradually decreases, which is again consistent with the continuity equation. The whole CIR and HSS structure (especially the compression region) is characterized by significantly increased magnetic field fluctuations that are mainly related to the Alfv\'enic-wave perturbations.

Transport of GCRs in the heliosphere is usually described in terms of the so-called Parker GCR transport equation \citep[][]{parker65}. It considers the GCR particle density at a given time and position, which is determined by the effect of the convection and diffusion (i.e., two terms from the standard Fokker-Planck equation) and by a term added by Parker, describing the effect of adiabatic cooling of particles within an expanding volume (or heating in the case of compression).
Later on, the Parker GCR transport equation was employed in many studies concerning various aspects of the heliospheric propagation and behavior of GCRs \citep[e.g.,][and very many references therein]{gleeson68,perko83,jokipii93,fisk98,potgieter98,wibberenz98,wibberenz00,morales09,morales10,gieseler16,zimbardo17,richardson18}.

The diffusion rate is determined by the diffusion coefficient, which depends on the mean-free path of particles, where the scattering is dominantly caused by the magnetic field fluctuations \citep[][]{parker65}.
While the diffusion coefficients are calculated, the corresponding total fluctuating magnetic energy is usually computed using the root mean square variation in the vector of the  interplanetary magnetic field \citep[e.g.,][]{zank98}.
The particle drift effect can also be included in the transport equation \citep[e.g.,][and references therein]{fisk98}, as well as the effect of mirroring on steep magnetic field gradients and shocks \citep[e.g.,][and references therein]{kirin20}. The diffusion regulates the inward transport of particles, whereas the solar wind flow regulates the outward transport of GCRs. The adiabatic term describes changes in the GCR energy spectrum.

This paper aims to explain the behavior of recurrent Forbush decreases caused by CIR and HSS structures in the solar wind at heliospheric distances about 1\,au, based on a convection-diffusion approach. Global 3D time-dependent simulations obtained by solving the GCR transport equation have shown that the smaller diffusion coefficient in the regions of stronger magnetic field related to CIRs results in a rapid local decrease in cosmic ray flux \citep[][]{kota91}. We also note that using the convection-diffusion approach to explain the CIR-related GCR modulation is not new \citep[e.g.,][]{chih86,wibberenz98}.
  \citet{richardson96} considered the steady-state diffusion-convection model, which includes adiabatic deceleration and longitudinal variations in the solar wind speed. In this model, the effect of the convection due to the outward flow of the solar wind is locally balanced by the inward diffusion of GCRs. The model that we present in this paper is to a certain degree similar to the more sophisticated 3D numerical model by \citet{alania11}. They regarded a steady-state heliosphere with CIR represented as the heliolongitudinal variation in solar wind speed and applied a full Parker transport equation. They found that the variation in GCR intensity is inversely correlated with the modulation parameter, which is proportional to the product of the solar wind velocity and the strength of the interplanetary magnetic field. Moreover, their calculations have shown that the heliospheric current sheet (HCS) structure in the model does not significantly affect the results. A similar result was also found by \citet{guo16}  using the 3D coupled MHD-CR transport model. They found that GCR variations depend on the ratio of diffusion coefficients in the fast and slow solar winds. Finally, we compare our results to the most recent dedicated 3D coupled MHD-CR transport models by \citet{kopp17} and  \citet{luo20}. Their simulations show that enhanced convection, reduced diffusion, and drifts due the magnetic field enhancement may contribute to the recurrent FD. However, the simulations cannot distinguish the relative contribution of different mechanisms.

In this respect we note that an application of the modeling results to observations and the corresponding quantitative analysis is still lacking. Hereafter, the GCR count rate measured in situ by spacecraft (s/c) is denoted as $CR$, whereas for the analog in the model option, the symbol $\Phi$ is applied.

For the purpose of our analysis, we use some observational examples related to a long-living CIR structure that was recorded over 27 solar rotations in the period from June 2007 to May 2009. It was analyzed in detail by
     \citet{dumbovic21}, hereafter, Paper 1; see also \citet{kuhl13}. In Section~\ref{cir} the observational aspect is described in detail, and in Section~\ref{model} the analytical model is proposed. A comparison of model results with observations is given in Sections~\ref{results} and \ref{analys}. The outcome of the analysis is discussed in Section~\ref{discus}, and conclusions are drawn in Section~\ref{concl}.

\section{CIR structure and the related GCR flux modulation}
\label{cir}

\begin{figure*}
\begin{center}
 \includegraphics[scale=0.9]{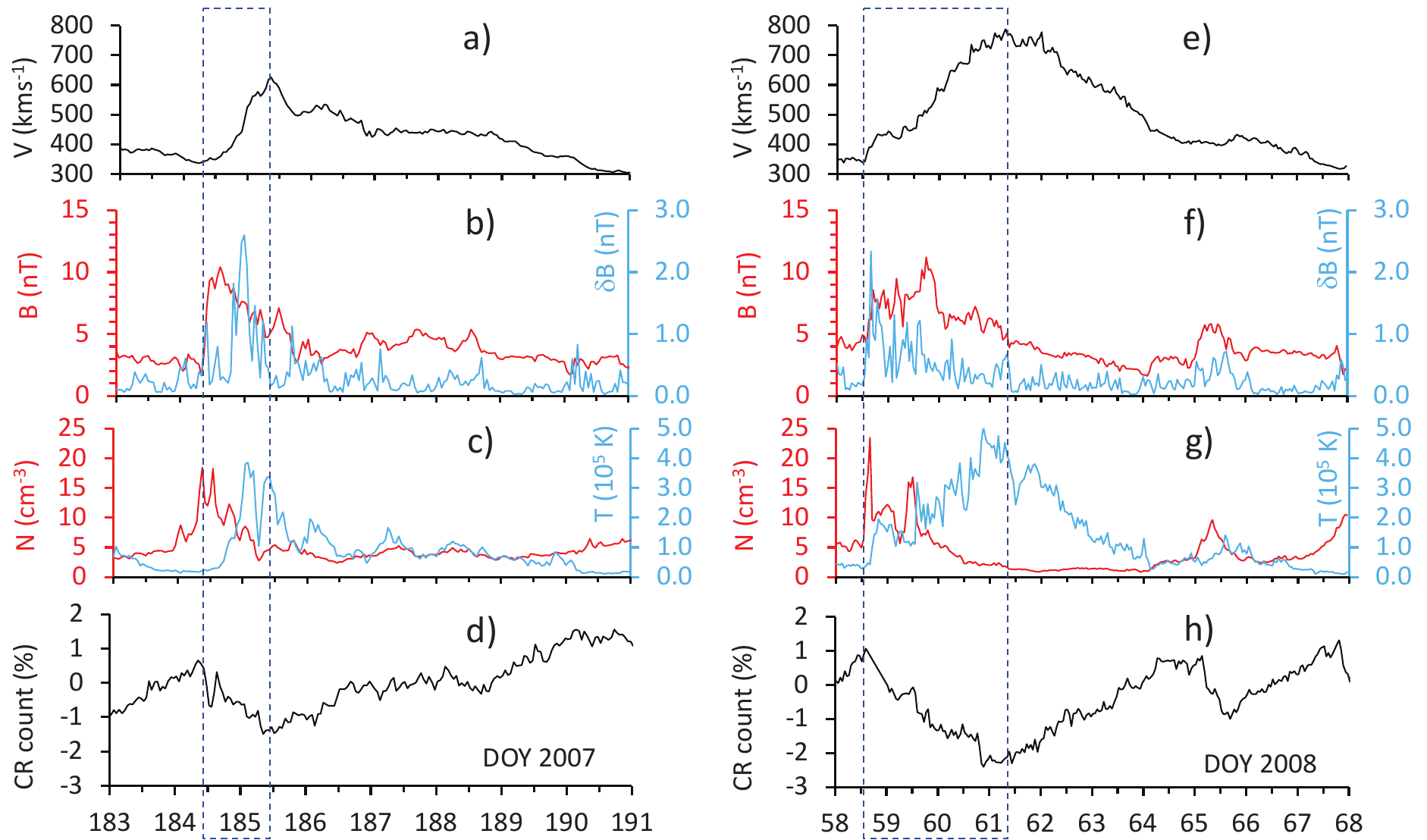}
\end{center}
\caption{
1\,au measurements of CIRs recorded from DOY 183 to 191, 2007 (rot-2; left column) and from DOY 58 to 68, 2008 (rot-11; right column). Panels a) and e): Flow speed, $V.$ Panels b) and f): Magnetic field strength, $B$ (red), and fluctuations
(root mean square),
$\delta B$ (blue). Panels c) and g): Proton number density, $N$ (red), and temperature, $T$ (blue). Panels d) and h): EPHIN cosmic ray count-rate change, $CR$, expressed in percent of the pre-event count rate. The dashed rectangle in both panels marks the period of the flow speed increase
}
\label{f1}
\end{figure*}

In Fig.~\ref{f1}, two examples of well-defined CIRs are presented. The events were recorded in the periods between the day of the year DOY\,=\,184 and 191, 2007 (Carrington rotation 2059; Fig.~\ref{f1}a), and DOY\,=\,58 and 68, 2008 (Carrington rotation 2067; Fig.~\ref{f1}b). Both CIR episodes were related to a long-living equatorial coronal hole that produced a persistent CIR and HSS  that lasted for 27 rotations. The presented examples were recorded during its second and eleventh rotation (hereafter, rot-2 and rot-11, respectively.
     In Figs.~\ref{f1}a--c  and e--g, the 1\,h resolution data from the OMNI database\footnote{https://omniweb.gsfc.nasa.gov} \citep[][]{king05} are displayed, and Figs.~\ref{f1}d and h show the 1\,h resolution data from the Electron Proton Helium
Instrument \citep[EPHIN;][]{muller95} on board the Solar and Heliospheric Observatory (SOHO), measured by its F-detector, which records hydrogen and helium nuclei above 53\,MeV \citep[][]{muller95}.
However, it is important to note that although the lower cutoff of SOHO-EPHIN-F is about 53 MeV, the response function of SOHO-EPHIN is such that it increases with energy, that is, it is less sensitive to low-energy particles \citep{dumbovic20}.

The rot-2 event began on DOY\,=\,184.3, 2007, starting from a quiet-Sun wind speed of $V\approx$\,340\,km\,s$^{-1}$ (Fig.~\ref{f1}a). The HSS speed attained a maximum of $\approx$\,610\,km\,s$^{-1}$ at DOY\,=\,185.3, after which it gradually decreased to only $\approx$\,305\,km\,s$^{-1}$ on DOY\,=\,190.8. The phase of the rising speed is marked in panels a-d of Fig.~\ref{f1} by the dashed blue rectangle. The period of increasing flow speed corresponds to the enhanced magnetic field, $B$, and plasma density, $N$ (red graphs in Figs.~\ref{f1}b and c, respectively) caused by the interaction of the HSS with the ambient slow solar wind. This period is also characterized by enhanced magnetic field fluctuations, $\delta B$, and rising temperature, $T$  (blue graphs in Figs.~\ref{f1}b and c, respectively). The change of the EPHIN cosmic ray count rate, $CR$, expressed in percentiles of the pre-event value, is presented in Fig.~\ref{f1}d. The $CR(t)$ graph very closely corresponds to the inverted $V(t)$ graph (for more examples, see Paper 1). The $CR$ data show a clear ``over-recovery'' phenomenon \citep[e.g.,][]{jamsen07,dumbovic12} after the main $CR$ dip.

The rot-11 CIR began on DOY\,=\,58.5, 2008, starting from a speed of $V\approx 350$\,km\,s$^{-1}$ (Fig.~\ref{f1}e). The HSS speed attained a maximum of $\approx$\,780\,km\,s$^{-1}$ at DOY\,=\,61.2, after which it gradually decreased to $\approx$\,320\,km\,s$^{-1}$ on DOY\,=\,68. The phase of the rising speed is marked in panels e-h of Fig.~\ref{f1} by the dashed blue rectangle. The period of increasing flow speed again corresponds to the enhanced magnetic field, $B$, and plasma density, $N$ (red graphs in Figs.~\ref{f1}f and g, respectively), as well as to the enhanced magnetic field fluctuations, $\delta B$, and rising temperature, $T$ (blue graphs in Figs.~\ref{f1}f and g, respectively). The change in the EPHIN cosmic ray count rate is presented in Fig.~\ref{f1}h. Again, the $CR(t)$ graph very closely corresponds to the inverted $V(t)$ graph. The data presented in Fig.~\ref{f1}f\,--\,g show a distinct secondary peak in the $B(t)$, $\delta B(t)$, and $N(t)$  graphs between DOY\,$\approx$\,65 and 66, corresponding to a weak increase in flow speed that is recognizable in the $V(t)$ graph (Fig.~\ref{f1}e), and is also notable as a weak new rise in the $T(t)$ graph (Fig.~\ref{f1}g). At the same time, $CR(t)$ shows a distinct secondary decrease, with a minimum at DOY\,=\,65.7 (Fig.~\ref{f1}h).

\begin{figure*}
\begin{center}
 \includegraphics[scale=0.7]{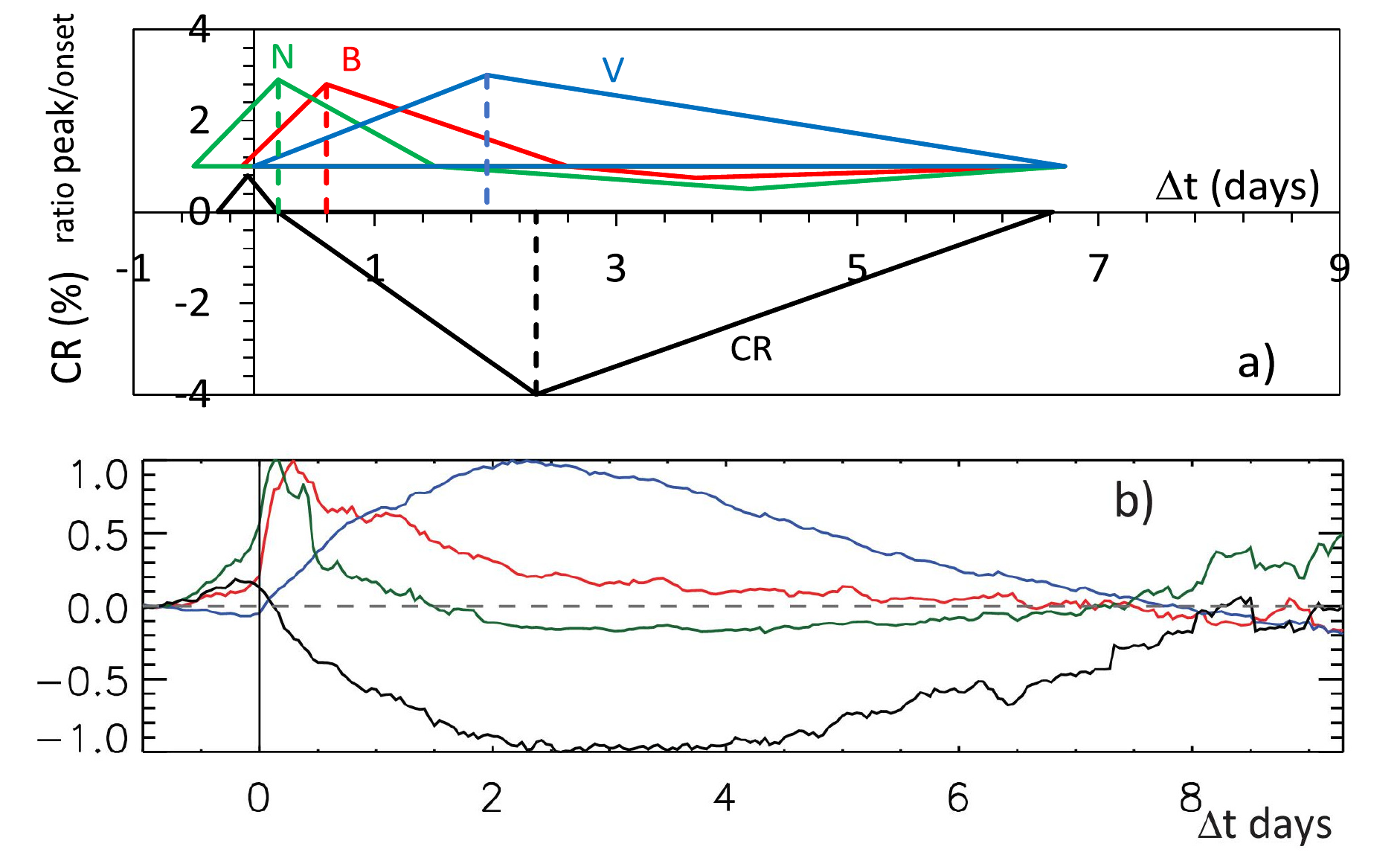}
\end{center}
\caption{
``Generic" temporal behavior of the flow speed, $V$ (blue), magnetic field, $B$ (red), density, $N$, (green), and the cosmic ray count-rate change expressed in percent, $CR$ (black), based on the study of a sample of CIRs presented in Paper 1.
 Panel a): Simplified, generic, presentation based on the mean values of the amplitudes and timing of the $B$, $N$, and $CR$ onsets, peaks, and ends, relative to the onset of $V$ (to see better the relative amplitude of $V$, it is multiplied by a factor of two). The peak times are marked by vertical dashed lines.
  Panel b): SEA results;  the variations relative to the pre-CIR values are normalized to the peak values, $\Delta V/V_{max}$, $\Delta B/B_{max}$, $\Delta N/N_{max}$, and $\Delta CR/CR_{max}$. The time $\Delta t=0$ is set at the onset of the $V$ increase.
}
\label{f2}
\end{figure*}

In Fig.~\ref{f2}a, a simplified generic temporal behavior of  CIR structure and the associated RFD is sketched schematically (based on the detailed statistical study presented in Paper 1).
The abscissa  shows the mean value of the relative time differences, $\Delta t$, of the onsets, maxima, and ends of $V$, $N$, $B$, and $CR$ relative to the onset of the flow-speed increase. It is expressed in days (thus, $\Delta t=0$ for the $V$ onset). The $y$-axis in the $y>0$ range represents the mean CIR-disturbance amplitudes, normalized with respect to the pre-CIR values, based on the mean values of the amplitudes of $N$, $B$, and $V$ (to highlight the relative amplitude of $V$, it is multiplied by a factor of two).
   The $[x_i,y_i]$ points obtained for the mentioned key parameters, that is,  onsets, maxima, and so on, are finally connected by straight lines (this is just a schematic drawing).
 The graph includes the rarefaction region at the rear of HSS. In the $y<0$ range, the mean RFD amplitude is shown as the percentage of the pre-CIR level.

The plasma density starts to increase before the onset of the flow-speed rise on average, whereas the magnetic field in this representation starts to increase approximately simultaneously with the flow-speed increase. After the compression period, a rarefaction phase occurs, characterized by weakening of the magnetic field and plasma density decrease \citep[e.g.,][and references therein]{hundhausen73,gosling96,gosling99,richardson18}. The $CR$ decrease occurs during the period of the speed increase, corresponding to the phase of the enhanced magnetic field. Before the $CR$ decrease, as is quite often observed, a  weak enhancement in the count rate is seen in the period when only the density is increased, like in the super-epoch analysis (SEA) representation (Fig.~\ref{f2}b). The $CR$ recovery phase closely follows the decrease in the flow speed.
   Regarding the SEA approach, it should be mentioned that \citet{badruddin16} analysed  the differences between characteristics of CIRs that developed a frontal shock and those without the shock. We emphasize that in our sample, CIRs did not show a shock signatures, and our SEA results are quite similar to those presented by \citet{badruddin16}.

Figure~\ref{f2}b shows an analogous, but more detailed, statistics-based generic time profile of $V$, $B$, $N$, and $CR$, where the displayed results are based on the SEA, performed in Paper 1, covering the complete set of 27 rotations. Unlike the schematic presentation shown in Fig.~\ref{f2}a, the graph displayed in Fig.~\ref{f2}b reveals that not only does $N$ starts to increase before $V$, but so does  $B$ (but within the standard deviation of $\Delta t$). Moreover, the graph $V(t)$ reveals a slight depletion before the main rise of $V$. The graph $CR(t)$ shows the pre-CIR slight increase in $CR$ in more detail \citep[e.g.,][]{lockwood71}, which is hereafter denoted as ``nose''. As we show below, the nose is most likely a consequence of the pre-event phase of the decreased $V$
and the steep magnetic field gradient related to the compression caused by the interaction of the fast and slow solar wind \citep[][]{kirin20}.

The characteristics  of the CIR structure and the associated RFD behavior presented in Figs.~\ref{f1} and \ref{f2} are the basis for constructing a model for the CIR-related RFDs. It is also used for a direct comparison of the observations and model results.

\section{Model}
\label{model}

\subsection{Physical background}
\label{basic}

\begin{figure*}
\begin{center}
\includegraphics*[scale=0.8]{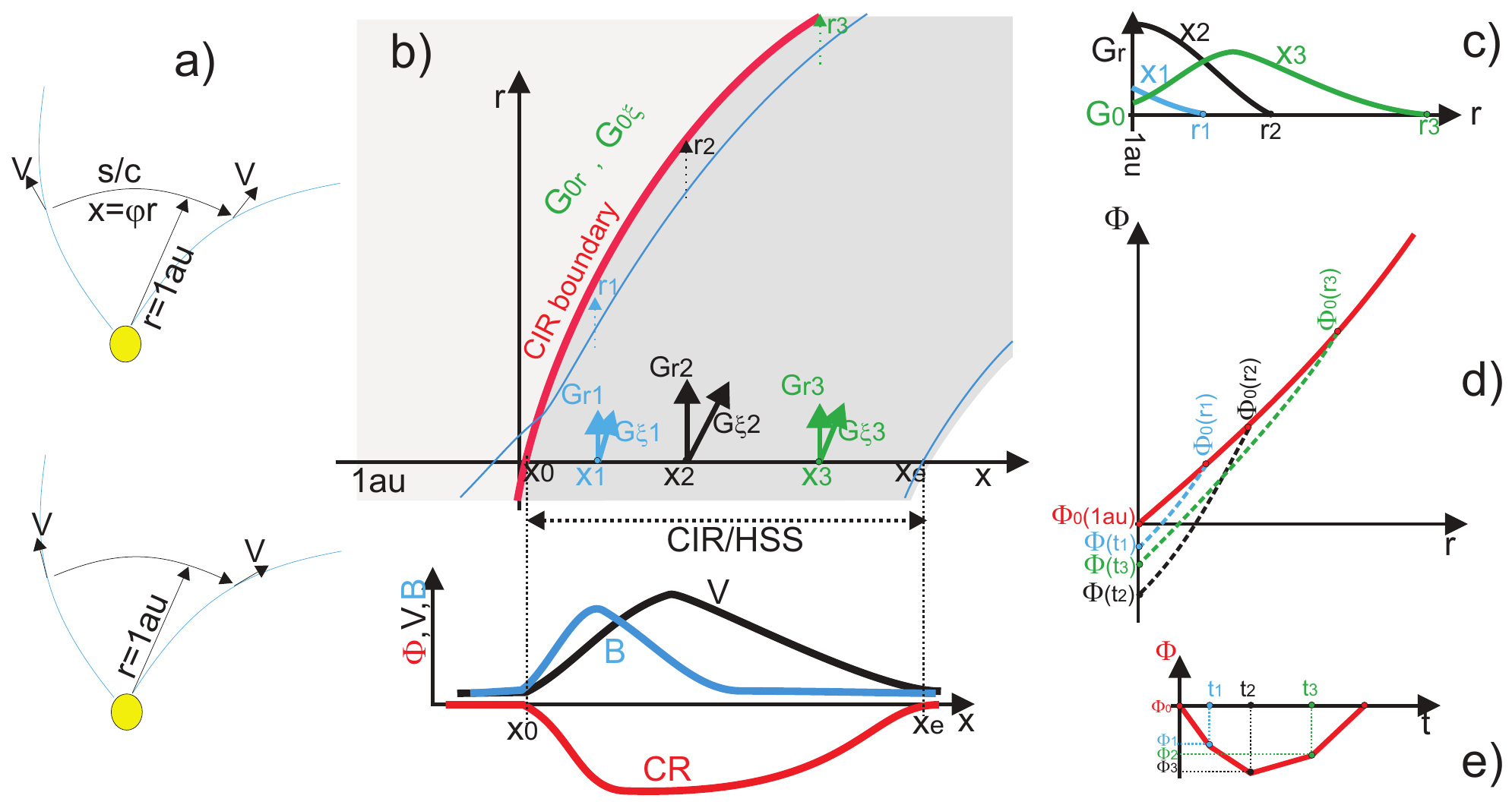}
\end{center}
\caption{
Panel a):
  Sketch of the global situation (upper panel: rest frame; lower panel: frame of reference rotating with the Sun); for details, see the main text. Panel b): Schematic presentation of the geometry and CIR characteristics
    (fixed reference frame),
 as applied in the model ($x_0$ and $x_e$ represent the beginning and end of CIR and HSS at 1\,au, respectively; for details, see the main text). Panel c): The dependence of the radial $CR$-gradient, $G_r$, for three specific s/c positions ($x_1$, $x_2$, $x_3$) as a function of heliocentric distance, $r$. The heliocentric distance of the CIR frontal edge for the three specific s/c positions, $x_i$, are denoted $r_1$, $r_2$, $r_3$. Panel d): The corresponding radial distance dependence of the cosmic ray count rate, $\Phi_{xi}(r)$, between CIR and HSS  outer boundary and s/c. The red line marks the behavior in a quiet-Sun situation at the CIR outer boundary. Panel e): Sketch of the resulting Forbush decrease profile, $\Phi(t)$, as measured by the s/c.
}
\label{f3}
\end{figure*}

In the following, we present a simple model for CIR and HSS effects on the cosmic ray count rate in the heliocentric distance range of $r=1$\,au (the global situation is sketched and the coordinate system explained in Figs.~\ref{f3}a and b). The model describes a single rotation, that is, rotations are considered independent  (differences from one to another rotation are due only to different quiet-Sun values of $\Phi_0$, $V_0$, and $B_0$, and the temporal behavior of $\Phi(t)$, $V(t)$, and $B(t)$ across the CIR and HSS at 1\,au).

The model is based on the Fokker-Planck diffusion equation,
\begin{equation}
\frac{\partial \Phi}{\partial t} + \nabla\cdot(\Phi {\bf V}) - \nabla\cdot (\kappa_{ij} \nabla \Phi ) = 0\,,
\label{e1}
\end{equation}
which was considered by \citet{parker65} as a staring point in developing a model for the heliospheric propagation of GCRs. In Eq.~(\ref{e1}), $\Phi$ represents the GCR particle number density at a given time and position, {\bf V} denotes the convective velocity, and $\kappa_{ij}$ represents the diffusion coefficient tensor. Eq.~(\ref{e1}) does not consider the change in particle energy spectrum due to the adiabatic heating/cooling of particles \citep[for a complete GCR transport equation, including the adiabatic effect, see][]{parker65}. In other words, we consider in the following only the change in total particle number density, regardless of the shape of the energy spectrum.

   We note that although the lower cutoff of SOHO-EPHIN is about 53 MeV, the response function of SOHO-EPHIN-F is such that it increases with energy, that is, it is less sensitive to low-energy particles. On the other hand, the spectrum of GCRs decreases with energy. As a result, the spectrum of particles detected by EPHIN therefore looks like a bell-like curve with a peak somewhere around 1 GeV \citep[depending on IP conditions, see, e.g.,][]{dumbovic20}.
  Bearing in mind that at 1\,au, the proton spectrum recorded by the EPHIN-F peaks somewhat below but close to 1\,GeV, small changes in proton energy spectrum therefore do not significantly affect the spectral distribution in the energy range $\approx$1\,GeV measured by s/c.

In the following, we consider a CIR structure in a stationary state, that is, we assume that its structure and characteristics do not change significantly over the time period required for its passage over the spacecraft. In this case, the time-derivative  $\partial \Phi/\partial t =0$ at a given position within the CIR and HSS, so that Eq.~(\ref{e1}) can be written as
\begin{equation}
\nabla\cdot(\Phi {\bf V}) = \nabla\cdot(\kappa_{ij} \nabla \Phi ) \,,
\label{e2}
\end{equation}
that is,
\begin{equation}
\Phi {\bf V} = \kappa_{ij} \nabla \Phi  \,,
\label{e3}
\end{equation}
which means that the effect of the convection due to the solar wind outward  flow is locally balanced by the inward diffusion of GCRs. The CIR-related decrease in $\Phi$ is a result of the suppressed  inward diffusion due to the enhanced magnetic field and its fluctuations, and it is also due to the enhanced outward convection. The diffusion coefficient parallel to the field lines $\kappa_{\parallel}$ is far larger at 1\,au than the perpendicular one $\kappa_{\perp}$, that is,  the diffusion perpendicular to the magnetic field is negligible compared to the diffusion throughout the field \citep[e.g.,][]{parker65,fisk98,giacalone98,richardson18}. We emphasize that the value of $\kappa_{\parallel}$ is quite ambiguous as it cannot be determined directly from observations. Its estimate therefore depends on a number of assumptions. Consequently, quite different values are found in various studies; they vary from $10^{17}$ to $>10^{19}$\,m$^2$\,s$^{-1}$. For example, \citet{parker65} estimated it to be $10^{17}-10^{18}$\,m$^2$\,s$^{-1}$,  $\approx3\times10^{19}$\,m$^2$\,s$^{-1}$ was reported by \citet{giacalone98}, $10^{18}$\,m$^2$\,s$^{-1}$ was adopted by \citet{gleeson68} and \citet{schwenn91}, and \citet{wibberenz98} considered values  $1.4-2.8\times10^{18}$\,m$^2$\,s$^{-1}$.
In the following, we use as a starting point for the quiet-Sun solar wind the radial diffusion coefficient of $\kappa_r=1.25\times10^{18}$\,m$^2$\,s$^{-1}$, which is obtained from the $CR$ radial gradient at 1\,au of $G_r=4.8$\,\%/au \citep[][]{gleeson68}  and for the solar wind speed of $V=400$\,km\,s$^{-1}$, and using $\kappa_r=V/G_r$ \citep[e.g.,][see also Eq.~\ref{e6}]{gleeson68,schwenn91}.

In constructing the model, we assumed that the diffusion coefficient is related to the magnetic field magnitude $B$, based on a number of previous studies \citep[e.g.,][and references therein]{potgieter13,richardson18}. In particular, we considered the simplest option, $\kappa\propto1/B$ \citep[see, e.g.,][]{potgieter98,wibberenz98,wibberenz00}. In addition, because the mean free path of particles depends on the magnetic field fluctuations \citep[e.g.,][and references therein]{wibberenz98,wibberenz00,michalek01,shalch04,chhiber17}, as does the diffusion coefficient, we also took the magnetic field fluctuations $\delta B$ into account, considering the simplest possible option   $\kappa\propto1/\delta B$.
    In this respect, we note that according to the model proposed by \citet{qin02}, which is based on standard quasi-linear theory, the parallel mean free path should be proportional to $B^{5/3}/\delta B^2$. As we show below, this is not consistent with observations, which show that the diffusion coefficient is roughly inversely proportional to both $B$ and $\delta B$.
Under these two approximations, the perturbed value of  the diffusion coefficient could therefore be estimated as $\kappa_B'=\kappa_0$\,$B_0/B$, or equivalently, $\kappa_{\delta B}'=\kappa_0$\,$\delta B_0/{\delta B}$, where $\kappa_0$, $B_0$, and $\delta B_0$ represent the values in the unperturbed solar wind.

However, the relation is most likely oversimplified, that is, it is more complex than the simplest options we assumed \citep[for other options, see, e.g.,][and references therein]{wibberenz98,wibberenz00,michalek01,shalch04,chhiber17}. For example, the mean free path, thus also the diffusion coefficient, depends on the particle energy, meaning that from the observational point of view, it depends on the energy channel of measurements (which is not of concern for our study because we considered only one channel). The relation $\kappa(\delta B)$ also depends on the spectrum of fluctuations. As a consequence, it probably depends on the heliocentric distance and on the solar cycle phase (in our case, this is not important as we considered a relatively short time period relative to the solar cycle duration, and we focused on heliocentric distances of about 1\,au). To compensate for all these ambiguities, we introduced a free parameter $k$ that is expected to be different from event to event due to differences in fluctuation time-spatial characteristics, and it would also depend on the energy channel of the measurements.

Thus, we introduce the scaling coefficients $k_b$ and $k_{\delta b}$, which we incorporated in $b^*=k_b$\,$b$ and $\delta b^*=k_{\delta b}$\,$\delta b$, where we used the abbreviations $b=B/B_0$ and $\delta b=\delta B/{\delta B_0}$. In other words, $b^*$ and $\delta b^*$ were used in the model calculations, where $k_b$ and $k_{\delta b}$ were determined by fitting the model results to the observations
  by minimizing the {\it rms} differences between the modeled and observed data.
 With this notation, the effective local radial gradient can be expressed as  $G^*_r=V/\kappa^*$, where $V$ is the local solar wind velocity, and $\kappa^*$ is defined by the unperturbed value of $\kappa_0$ and the effect of the perturbed magnetic field, as determined by the parameters $b^*$ or $\delta b^*$, that is, $\kappa^*=\kappa_0/b^*$ or $\kappa^*=\kappa_0/\delta b^*$, which means $G^*=(V/\kappa_0)b^*=G_0b^*$ and equivalently, $G^*=(V/\kappa_0)\delta b^*=G_0\delta b^*$.

We now consider as the first step of the following considerations the situation in the reference frame that rotates with the Sun, that is, the frame whitin which the CIR and HSS are at rest, whereas the s/c passes through it. In this reference frame, the solar wind flow velocity, which is radial in the rest frame of reference, becomes aligned with the magnetic field lines because the velocity now has an additional component, corresponding to the solar rotational velocity \citep[see Fig.~\ref{f3}a, and, e.g., Figs. 1, 4, and 5 in][]{hundhausen72}. The advantage of the frame rotating with the Sun is that the diffusion and the wind flow are coaligned, both following the magnetic field lines, so that the diffusion is antiparallel with the convective flow. Because $\kappa_{\perp}\ll\kappa_{\parallel}$ means that the distribution of $\Phi$ along a given field line, $\Phi(\xi)$, where $\xi$ represents the curvilinear coordinate along the field line, is independent of the distribution of $\Phi$ along neighboring field lines. Thus, considering the situation along a particular field line, Eq.~(\ref{e3}) can be written as
\begin{equation}
\Phi(\xi) V_\xi = \kappa_\parallel\,\frac{\partial \Phi(\xi)}{\partial\xi} \,,
\label{e4}
\end{equation}
that is,
\begin{equation}
\frac{\partial (\ln\Phi(\xi))}{\partial\xi} = G_\xi\,,
\label{e5}
\end{equation}
where we have introduced the abbreviation for the GCR gradient along a given field line,
\begin{equation}
G_\xi = \frac{V_\xi}{\kappa_\parallel}  \,.
\label{e6}
\end{equation}
 Eq.~(\ref{e6}) defines in firm physical terms a local gradient of $\Phi(\xi)$, $G_\xi$, along a given field line. If $G_\xi$ is approximately constant over a distance range from $\xi_1$ to $\xi_2$ along a particular field line, or in other words, if we take the mean value of $G_\xi$ over this segment of the field line, $\overline G_\xi$, the solution of Eq.~(\ref{e5}) reads
\begin{equation}
\Phi(\xi_2) = \Phi(\xi_1)\,\,e^{\overline G_\xi (\xi_2-\xi_1)}\,.
\label{e7}
\end{equation}
For the vicinity of 1\,au, we can write
\begin{equation}
\Phi(\xi) = \Phi_{1\rm{au}}\,e^{\overline G_\xi \xi}\,,
\label{e8}
\end{equation}
where $\Phi_{1\rm{au}}$ represents the value of $\Phi$ at s/c located at 1\,au, and we set $\xi=0$ at the intersection of a given field line with the 1\,au s/c trajectory.

This means that when we have the 1\,au s/c measurements of $\Phi_{\rm 1au}$ (or analogously, at any other heliocentric distance at which a particular s/c is located), as well as the 1\,au measurements of $V$ and $B$ (providing an estimate of $G_\xi$), Eq.~(\ref{e8}) provides an estimate of $\Phi$ at any position [$x,r$] within or out of CIR and HSS. Moreover, if measurements of $\Phi(t)$ and $V$ from two s/c located at two different heliocentric distances, $r_1$ and $r_2$, were available, we could easily estimate the value of $\overline G_\xi$ within the corresponding distance range $\xi_1$\,--\,$\xi_2$ along a particular field line, assuming the nominal shape of the corresponding Parker spiral. This would also provide an estimate of the mean value $\overline\kappa_\parallel$ within the distance range $\xi_1$\,--\,$\xi_2$ by using Eq.~(\ref{e6}). By also employing the measurements of $B$ and $\delta B$, we can also find an empirical expression relating $\kappa_\parallel$ to $B$ or $\delta B$.

Finally, we emphasize that Eq.~(\ref{e8}) clearly shows that after the s/c exits the HSS, in an ideal case in which the flow speed and the magnetic field attain pre-CIR values, that is, the value of $G_\xi$ returns to the initial state, the cosmic ray particle density should also return to the pre-event level. If the flow speed and the magnetic field values differ from those measured before the CIR, the post-event cosmic ray count rate level should be different from that recorded in the pre-event state. If values of $V$ and $B$ are lower than in the pre-event state, the so-called over-recovery would be observed, in which the cosmic ray count rate after the HSS is higher than prior to it. If values of $V$ and $B$ in the post-HSS phase remain higher than in the pre-event phase, the count-rate will stay lower than it was before the CIR.

\subsection{``Operative'' approximation}
\label {oper}

Although the consideration presented above describes the basic behavior of $\Phi$ throughout the CIR and HSS structure, it is inconvenient for quantitative analysis and comparison with the measured values of $CR$ at 1\,au because the values of $\Phi_{1\rm{au}}$ at different 1\,au s/c positions that are basically governed by Eq.~(\ref{e5})  cannot be related directly to the pre-event values $\Phi_0$. This is primarily
because in the absence of measurements by two (or more) s/c overtaken by the same CIR structure at different heliocentric distances, the value of $\Phi$ at a certain referent distance is not available \citep[for uncertainties of $\Phi$ at large heliocentric distances, see, e.g., Fig. 3 in][]{mckibben82}.

In the following, we therefore approach the problem in an alternative way that provides an approximative, but directly operative quantitative analysis of $\Phi_{1{\rm au}}(t)$. More precisely, the proposed procedure described below enables estimating the ratio of the value of $\Phi$ at 1\,au along the $x$-axis (for the definition of the coordinate system we used, see Fig.~\ref{f3};
   in Fig.~\ref{f3}b we show the situation in the fixed reference frame)
relative to the pre-event value $\Phi_0$, that is, the values of $\Phi_{1{\rm au}}(x)/\Phi_0$, which can then be readily converted into the ratio $\Phi_{1{\rm au}}(t)/\Phi_0$, as measured by s/c during the CIR and HSS passage over it. To do this, we converted the dependence $\Phi_x(\xi$)  into an equivalent that defines the dependence of $\Phi$ on the radial coordinate, $r$, at a given $x$-coordinate, $\Phi_x(r)$. In other words, the gradient $G_{\xi}(x)$ as estimated for 1\,au was converted into the radial gradient, $\overline G_r(x)$.

The procedure of obtaining the ratio  $\Phi_{1{\rm au}}(x)/\Phi_0$ was started by calculating the values of $G_\xi$ along the $x$-axis $G_\xi(x,1{\rm au})$. Then, taking into account the nominal orientation of the local Parker spiral
   (geometrically, the Archimedian spiral),
we calculated the component of the gradient $G_\xi(x,1{\rm au})$ in $r$-direction to obtain local values, $G_r(x,1{\rm au})$.
   The Archimedian spiral is defined in polar coordinates [$r,\theta$] as
\begin{equation}
 r= \frac{v_r}{\omega} \,\theta\,,
\label{archimed}
\end{equation}
   where $v_r$ is a constant radial speed and $\omega$ is a constant rotational speed. The winding angle, representing the tilt angle of the field lines relative to the radial direction, reads
\begin{equation}
\tan \psi= \frac{B_\phi}{B_r}=\frac{\omega r}{v_r }\,.
\label{tilt}
\end{equation}
 The winding angle relates $G_\xi$ and $G_r$ as
\begin{equation}
G_r=G_\xi\tan \psi\,.
\label{gg}
\end{equation}

This situation is sketched in Fig.~\ref{f3}b (fixed reference frame) for three different s/c positions, $x_i$, denoted as $x_1$, $x_2$, and $x_3$, where the corresponding values $G_{\xi 1}$, $G_{\xi 2}$, and $G_{\xi 3}$ are also sketched together with the related components $G_{r1}$, $G_{r2}$, and $G_{r3}$. The CIR and HSS structure is shaded in gray, and the frontal edge of the CIR is marked by the bold red line. The magnetic field lines are sketched by the thin blue lines. The heliocentric distance of the CIR boundary at coordinates $x_1$, $x_2$, and $x_3$ are denoted as $r_1$, $r_2$, and $r_3$, respectively. Ahead of the CIR boundary lies a quiet-state slow solar wind (shaded light gray), characterized by the GCR radial gradient $G_{r0}$ and the field-aligned gradient $G_{\xi0}$. At the bottom of Fig.~\ref{f3}b, the typical behavior of the 1\,au solar wind flow speed, $V$, the magnetic field magnitude, $B$, and the cosmic rate count rate, $CR$, are sketched schematically.

In the next step, we estimated for each s/c position $x$ within CIR and HSS the values of $G_r(x,r)$ along the $r$-direction within the CIR and HSS structure, that is, from $r=1$\,au up to the outer CIR boundary located at $r=r_{\rm cir}(x)$, meaning that we converted the measurements obtained by s/c at 1\,au into what is expected in the entire  CIR and HSS structure within the range from its 1\,au onset to its 1\,au end. To illustrate this mapping procedure, we considered a moment $t=t_i$ after $t=0$, set at the CIR onset as measured by the s/c. At this moment, the plasma element that was at 1\,au at $t=0$ now already lies at $r_{\rm cir}=1{\rm au}+V_{r0}$\,$t_i$, where $V_{r0}$ is the solar wind velocity measured at $t=0$, so that $r_{\rm cir}$ represents the radial distance of the frontal CIR boundary at $t=t_i$. Analogously, we can track the trajectory of any plasma element that was at 1\,au in the period from $t=0$ to $t=t_i$. For example, the element that was at 1\,au at $t_j<t_i$, showing the radial speed $V_{j}$, at $t_i$ will be at the distance $r_{j}(t_i)=1{\rm au}+V_{j}$\,$(t_i-t_j)$. When we take into account that the solar wind speed around 1\,au does not change with distance, and when we also assume that in the considered distance range the magnetic field does not change much, the physical conditions within a given plasma element remain mainly unchanged over its trajectory within the $t_i-t_j$ interval. Thus, the  values of $G_r$ estimated from s/c measurements (as previously described) for any $t_j$ in the period from $t=0$ to $t_i$ can be attributed to the corresponding distance $r_{j}$ between $r_{\rm cir}(x_i)$ and 1\,au. Based on this mapping of $G_r(r)$ at each $x_i$, it is straightforward to estimate the mean value of $G_r$ from 1\,au to $r_{\rm cir}$ for a given s/c position $x_i$, that is, for the corresponding moment $t_i$.

As a digression, we note here that the increasing $V_r$ with the increasing time $t_i$, that is, increasing distance $x_i$ within the CIR and HSS structure, also means that faster plasma elements catch up the preceding slower elements. Thus, over a larger heliospheric distance range, the previous approximations cannot be applied because the plasma and magnetic field within a plasma element are compressed due to $[\partial V_r/\partial r]_x<0$.
However, the compression effect is partly compensated for by the $1/r^2$ lateral expansion of the structure.

The steepening of the frontal CIR structure leads to one more important effect, and that is the forward-shock formation \citep[][]{gosling96,gosling99}. As the spatial $V_r(r)$ profile causes a gradual steepening of the frontal part of the CIR, it will at a certain distance $r_s$ inevitably transform into a shock. Denoting the radial speed of the frontal CIR element as $V_{r0}$, the maximum 1\,au HSS radial speed as $V_r^{\rm max}$, their difference as $V_r^{\rm max}-V_{r0}\equiv \Delta V$, the time elapsed from the 1\,au CIR onset to the moment at which the HSS attains $V_r^{\rm max}$ as $\Delta t_{\rm max}$, and the corresponding distance between the CIR outer boundary and 1\,au as $\Delta r$ , we can estimate the time needed for the shock formation as $\Delta t=\Delta r/\Delta V=V_{r0}\Delta t_{\rm max}/\Delta V$.

We take as typical examples the measurements presented in Figs.~\ref{f1}a and e, where the frontal radial speed was $V_{r0}\approx 350$\,km\,s$^{-1}$ and the maximum speeds within the HSS were $V_r^{max}\approx 600$ and 800\,km\,s$^{-1}$, achieved within $\approx1$ and 3\,d, respectively. The corresponding times needed for the full completion of the shock after the frontal CIR edge passed over the 1\,au s/c are then equal to 1.4 and 2.3\,d, respectively, and the corresponding heliocentric distances  are $\approx 1.5$ and 2.1\,au. This estimate is broadly consistent with observations as well as with numerical results \citep[e.g.,][]{gosling96,gosling99}.

We return to the previous considerations of the radial GCR gradient. The obtained values of $G_r(x_i,1{\rm au})$, whose behavior is schematically sketched in Fig.~\ref{f3}c for three different s/c positions, $x_1$, $x_2$, and $x_3$, provide the estimate of the cosmic ray count rate variation along $r$ for a given s/c position $x_i$. Beyond the outer CIR boundary, located at $r_{\rm cir}(x_i)$ the value of $G_r(x_i, r_{\rm cir}(x_i))$ is taken to be constant \citep[e.g.,][]{mcdonald81}, having the value $G_{0r}$ (Fig.~\ref{f3}b), which can be chosen from some of the values reported in a number of previous studies, for instance,  4.8\,\%/au, as reported by \citet{gleeson68};  3.5\,\%/au, as estimated by \citet{mcdonald81}; 1\,--\,10\,\%/au, and 3\,--\,7\,\%/au over different phases of the solar cycle, as reported by \citet{fisk98} and \citet{fujii93}, respectively; 2.7\,\%/au, as found by \citet{desimone11}; and \citet{gieseler16};  2\,--\,6.6\,\%/au, as determined for different energy ranges by \citet{marquardt19}. These values depend not only to the phase of the solar cycle and the heliospheric distances considered, but also on the proton energies taken into account, consistent with the predictions by \citet{parker65}.

 This enables an estimate of the values of $\Phi(x,1{\rm au})$, starting from the quiet-state value of $\Phi(x,r_{\rm cir})$ at the outer boundary of CIR and HSS as defined by the quiet-state radial gradient $G_{0r}$. The behavior of $\Phi_{xi}(r)$ at a particular $x$-coordinate, $x_i$, is now described by the equation for the $r$-direction,
\begin{equation}
\frac{1}{\Phi_{xi}(r)}\frac{\partial \Phi_{xi}(r)}{\partial r} = G_r(x_i,r)\,.
\label{e9}
\end{equation}
 Eq.~(\ref{e9}) defines the behavior not only at 1\,au, but rather, at any heliocentric distance $r$ within, as well as beyond 1\,au, as given by the solution of Eq.~(\ref{e9}),
\begin{equation}
\ln{\Phi_{xi}(r)} = \int G_r(x_i,r)\,{\rm d}r\,.
\label{e10}
\end{equation}
To obtain the values of $\Phi$ at 1\,au for a given $x_i$ coordinate, Eq.~(\ref{e10}) has to be integrated step by step, starting from $r_{\rm cir}(x_i)$, where the the cosmic ray count rate has the nominal quiet-state value for that heliospheric distance, $\Phi_0(r_{\rm cir}(x_i))$, down to $r=1$\,au to derive the needed value $\Phi_{1\rm au}(x_i)$.

The heliocentric distance
\begin{equation}
r_{\rm cir}(x_i) = 1{\rm au}+\Delta r_{\rm cir}(x_i)\,
\label{e11}
\end{equation}
is calculated by substituting into Eq.~(\ref{e11}) the value $\Delta r_{\rm cir}(x_i) = V_{0r}\Delta t_i$, where $V_{0r}$ denotes the pre-event radial solar wind speed, and $\Delta t_i$ represents the time measured from the onset of CIR at 1\,au, $t_0$, that is., $\Delta t_i = t_i - t_0$.

The obtained values of $G_r(x,r)$ within the CIR and HSS structure define the behavior of $G_r(r)$ for any coordinate $x_i$. This is schematically sketched in Fig.~\ref{f3}c for three different $x$-coordinates, $x_1$, $x_2$, and $x_3$, corresponding to segments at the beginning portions of the CIR, the place where $G_r(1\,{\rm au})$ is maximum, and a portion within the rear parts of HSS, respectively. Thus, Eq.~(\ref{e10}) can now be evaluated to obtain $\Phi_{xi}(r)$, and thus $\Phi_{xi}(1{\rm au})$, and consequently, $\Phi_{1{\rm au}}(t_i)$. This is schematically depicted in Figs.~\ref{f3}d and e, respectively.

On the one hand, bearing in mind all the approximations that we used, and on the other hand, the inherent noise in measurements and the process of smoothing, we simplify below the calculation of $\Phi_{xi}(1{\rm au})$, i.e., $\Phi_{1{\rm au}}(t_i)$, where instead of integrating Eq.~(\ref{e10}) step by step along the $r$-direction at a given $x$-coordinate, $x_i$, we estimate the mean value of $G_r$, along $r$ at $x_i$, to get $\overline G_r(x_i)$, and use it to obtain the value of $\Phi_{1{\rm au}}(x_i)$ as
\begin{equation}
\Phi_{1{\rm au}}(x_i) = \Phi(r_{{\rm cir}}(x_i))\,\,{\rm e}^{-\overline G_r(x_i)\Delta r(x_i)} \,,
\label{e12}
\end{equation}
where $\Delta r(x_i)=r_{\rm cir}(x_i)-1$au represents the radial extent of CIR and HSS beyond 1\,au at a given s/c position $x_i$, as defined below Eq.~(\ref{e11}).

Having the values of $\Phi_{1{\rm au}}(x_i)$, i.e.,  $\Phi_{1{\rm au}}(t_i)$, the values of $\Phi_{1{\rm au}}(t_i)$ can be straightforwardly converted into the change of $\Phi_{1{\rm au}}$ relative to the pre-CIR value $\Phi_0$, expressed in percentiles,
\begin{equation}
\Delta\Phi_\%(t) = 100\,\, \frac{\Phi_{1{\rm au}}(t) - \Phi_0}{\Phi_0} =  100\, \left(\frac { \Phi_{1{\rm au}} (t)} {\Phi_0}  - 1\right) \,,
\label{e12}
\end{equation}
which can be directly compared with the observations in the form as presented in Figs.~\ref{f1}d and h, or the statistics-based generic form shown in Figs.~\ref{f2}a and b.

Finally, we note that the described model is to a certain degree similar to the propagating diffusive barrier approach \citep[PDB,][]{wibberenz98,wibberenz00}, which describes the effects of merged interaction regions (MIRs) at large heliocentric distances. However, we note that in PDB, which are valid for large heliocentric distances at which  the magnetic field lines are almost concentric circles, the radially outward convection is balanced by the diffusion perpendicular to the magnetic field.

\section{Comparison with observations}
\label{results}

In the following, the model proposed in the previous section is applied to the observations, using the results of the study presented in Paper 1, where the characteristics of a long-living CIR that was recorded over 27 solar rotations in the period from June 2007 to May 2009, are analyzed in detail. First, we consider in Section~\ref{sea} the generic scheme shown in Fig.~\ref{f2}a, as well as the results of the SEA displayed in Fig.~\ref{f2}b. Then, in Section~\ref{rot} we consider two specific examples, recorded from DOY 183 to 191, 2007 (rot-2) and from DOY 58 to 68, 2008 (rot-11). They are presented in Fig.~\ref{f1}.

As specified in Section~\ref{model}, the main model input parameters are the solar wind radial flow speed, $V_r$ and the magnetic field strength, $B$, or its fluctuations, $\delta B$. We also note that we employed the previously mentioned free parameter $k$ when we relate the magnetic field strength (or its fluctuations) to the radial gradient of the GCR count rate, $G_r$, because the dependence of the diffusion coefficient on the magnetic field strength is not well known and the background heliospheric GCR radial gradient, $G_{r0}$, as  inferred in a number of previous studies, spans quite a broad range of values. The GCR gradient should be higher in stronger fields, that is, roughly $G_r\propto B$ due to $\kappa_\parallel \propto 1/B$ \citep[][]{potgieter98,wibberenz98,wibberenz00}. We emphasize that the free parameter $k$ does not affect the overall behavior of the modeled cosmic ray count rate within CIR and HSS. It is only used to adjust the overall amplitude of the model outcome to the observed amplitudes of the cosmic ray count rate variations. In other words, when the value of $k$ is determined from fitting the model results to the observations, we obtain a better insight into the relation between the diffusion coefficient and the magnetic field strength, or its fluctuations.

\subsection{Generic SEA and ``schematic'' time profiles}
\label{sea}

\begin{figure*}
\begin{center}
\includegraphics[scale=0.90]{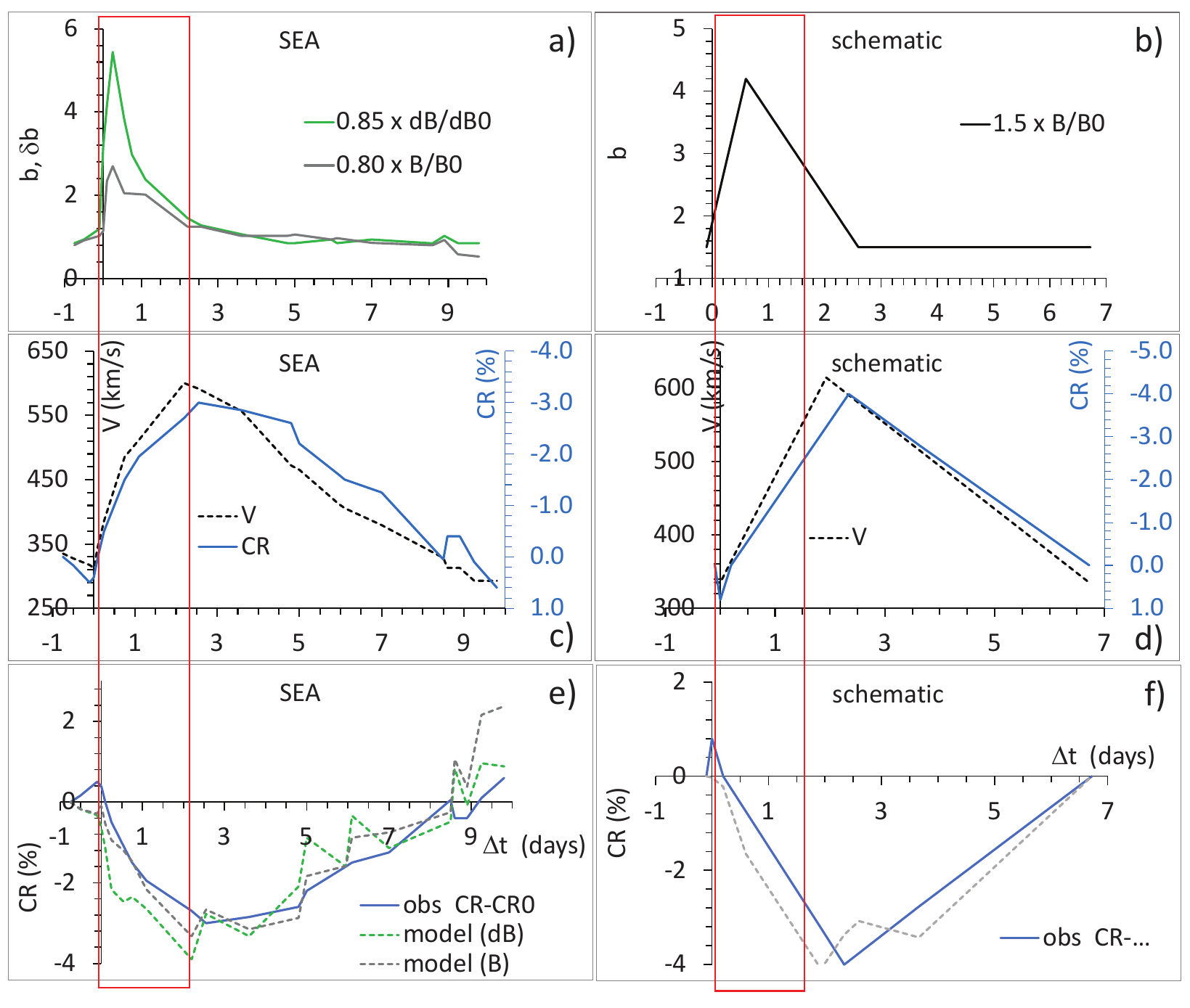}
\end{center}
\caption{
Comparison of the model results and observations: SEA, based on the study in Paper 1, shown in Fig.~\ref{f2}a  (left) and the schematic generic behavior, based on Fig.~\ref{f2}b (right). Panel a):  Magnetic field, $B$ (gray), and its fluctuations $\delta B$ (green), normalized with respect to the pre-event values, $b=B/B_0$ and $\delta b=\delta B/\delta B_0$, multiplied by the free parameter $k$ (written in the inset), based on SEA. Panel b):  Magnetic field, $B,$ normalized with respect to the pre-event values, $b=B/B_0$, multiplied by the free parameter $k$ (written in the inset), based on the schematic behavior. Panels c) and d): Solar-wind flow speed, $V_r$ (dashed black line) and the inverted graph of the cosmic ray count rate, $CR$ (blue line). Panels e) and f): observed $CR$ (blue line) compared to the model results (dashed gray line) based on the behavior of $b$; in Panel e) model results for the $\delta b$ option are presented as the dashed green line.
 The red rectangles mark the period of the rising solar wind speed.
}
\label{f4}
\end{figure*}

In the left column of Fig.~\ref{f4} we show a comparison of the outcome of SEA (see Fig.~\ref{f2}b), with the model result for scaling factors $k_B = 0.8$ and $k_{\delta B}=0.85$, estimated by matching the model with the observations.
  The values of coefficients $k$ were obtained by minimizing the {\it rms} differences between the modeled and observed data.
The model output is based on a smoothed $V_r(t)$ curve and smoothed curves $b(t)=B(t)/B_0$ and $\delta b(t)=\delta B(t)/\delta B_0$.
   The smoothing is based on segmental linear fitting (hereafter, SLF), where linear fitting is applied to each segment between two neighboring knees in the measured time curves of a given parameter
   (for details, see Appendix A). We chose the SLF option instead of directly comparing the 1 h observed data with the model results simply to reduce the noise in the observed-minus-model data, especially in the cases of $B$ and $\delta B,$ which are characterized by quite high noise.
In Fig.~\ref{f4}c the SEA SLF-smoothed $V_r(t)$ curve (dashed black line) is compared with the inverted curve of the SEA SLF-smoothed cosmic ray count rate variation, $CR(t)$ (blue lines), to illustrate a close similarity of the $V_r(t)$ and $-CR(t)$ behavior. In Fig.~\ref{f4}e the modeled SEA-based $\Phi(t)$ curves are directly compared to the SLF-smoothed SEA $CR(t)$ curve. The red rectangle emphasizes the period of rising solar wind speed, $\partial V_r/\partial t>0$.

In Figs. ~\ref{f4}b, d, and f, a comparison of the generic schematic time profiles (Fig.~\ref{f2}a) and the model $B$ option results for $k=1.5$  is presented, analogously to that concerning the generic SEA-option shown in Figs.~\ref{f4}a, c and e, respectively. In both options, model results and measured data match well. Moreover, the nose in the $CR$ time profile corresponds to a slight pre-CIR depletion in the velocity curve.

\subsection{Two specific examples: rot-2 and rot-11}
\label{rot}

\begin{figure*}
\begin{center}
\includegraphics[scale=0.9]{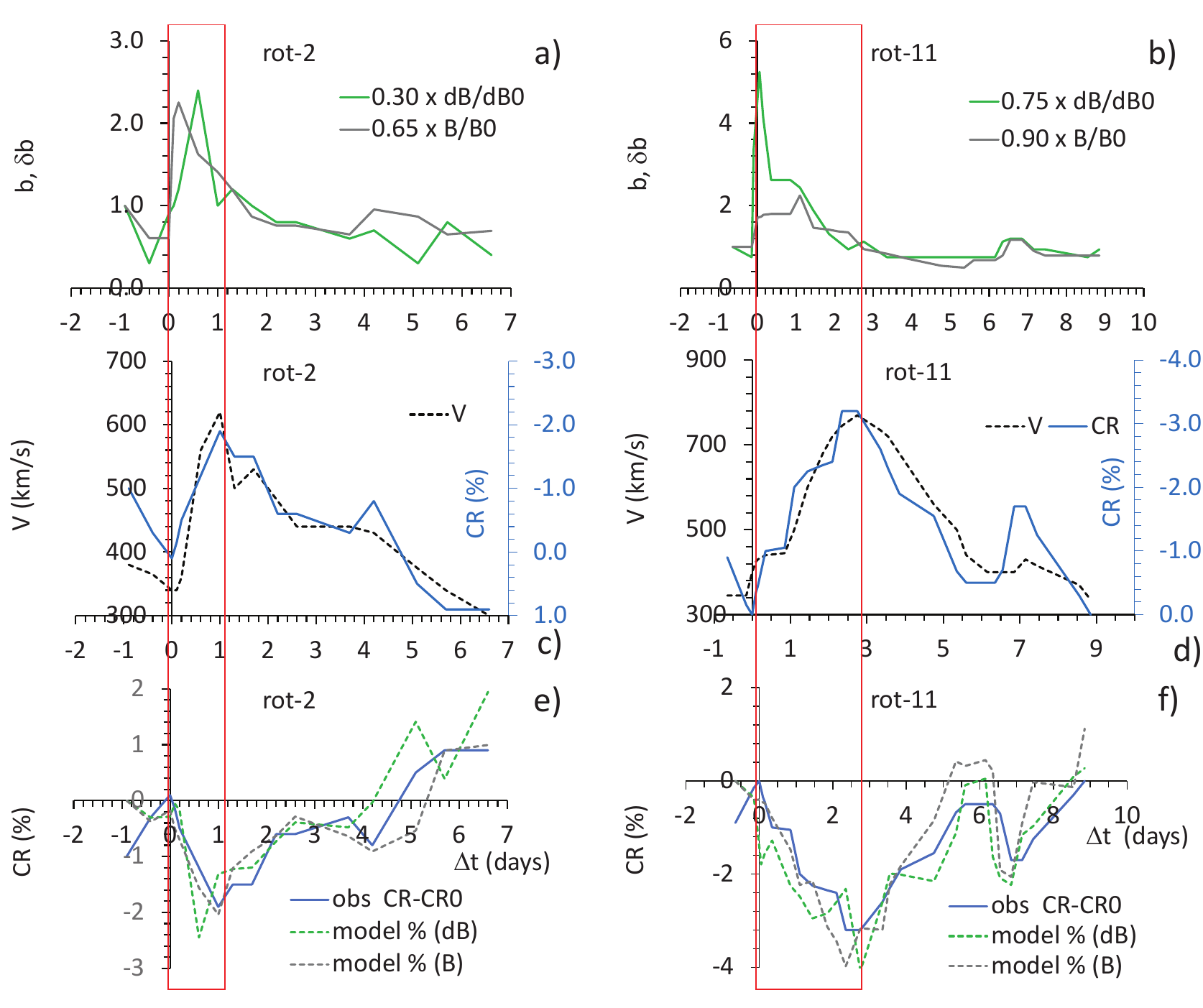}
\end{center}
\caption{
Comparison of the model results and observations for two CIR events. Left: rot-2 (DOY\,=\,184 and 191, 2007). Right: rot-11 (DOY\,=\,58\,--\,68, 2008). Panels a) and b): magnetic field and its fluctuations normalized with respect to the pre-event values ($b$ and $\delta b$, gray and green; respectively). Panels c) and d): solar-wind flow speed (dashed black) and the inverted $CR(t)$ graph (blue). e) and f) the observed $CR(t)$ values (blue) compared to the model results based on the behavior of $b$ (dashed gray) and $\delta b$ (dashed green).
}
\label{f5}
\end{figure*}

In Fig.~\ref{f5} we present a comparison of the model results and observations for rot-2 (left column) and rot-11 (right column). These two examples are chosen due to their well-defined and ``clean'' CIR and HSS structure, that is, there was no effect of other transient disturbances, such as ICMEs or another CIR that overlapped those we studied. Furthermore, they are different to a certain degree, not only by the amplitudes and perturbation duration, but also because rot-2 shows the over-recovery of $CR$ at the end of HSS (Fig.~\ref{f5}e, see also Fig.~\ref{f1}d), whereas rot-11 shows a distinct secondary peak in $B$ and $\delta B$ at the rear of HSS with a corresponding secondary decrease of $CR$ (Fig.~\ref{f5}f, see also Fig.~\ref{f1}h).

Figures~\ref{f5}a and b show the SLF-smoothed behavior of the normalized magnetic field strength, $b=B/B_0$, and its fluctuations, $\delta b=\delta B/\delta B_0$, for rot-2 and rot-11, respectively.  The displayed values are scaled by the factor $k$ we used in the calculations ($k_b=0.65$ and 0.9 for $b$, and $k_{\delta b}=0.3$ and 0.75 for $\delta b$, respectively). In Figs.~\ref{f5}c and d, the SLF-smoothed behavior of $V_r$ is drawn together with the SLF-smoothed inverted $CR$ curve to emphasize the similarity of the two curves. In Figs.~\ref{f5}e and f, the observed behavior of $CR$ (blue) is compared with the model results based on $b(t)$ (dashed gray) and $\delta b(t)$ (dashed green). The graphs are very similar to the modeled and observed behavior, including the over-recovery in the case of rot-2, as well as the secondary dip in rot-11.

\section{Analysis of the model results}
\label{analys}

\begin{figure*}
\begin{center}
\includegraphics[scale=0.90]{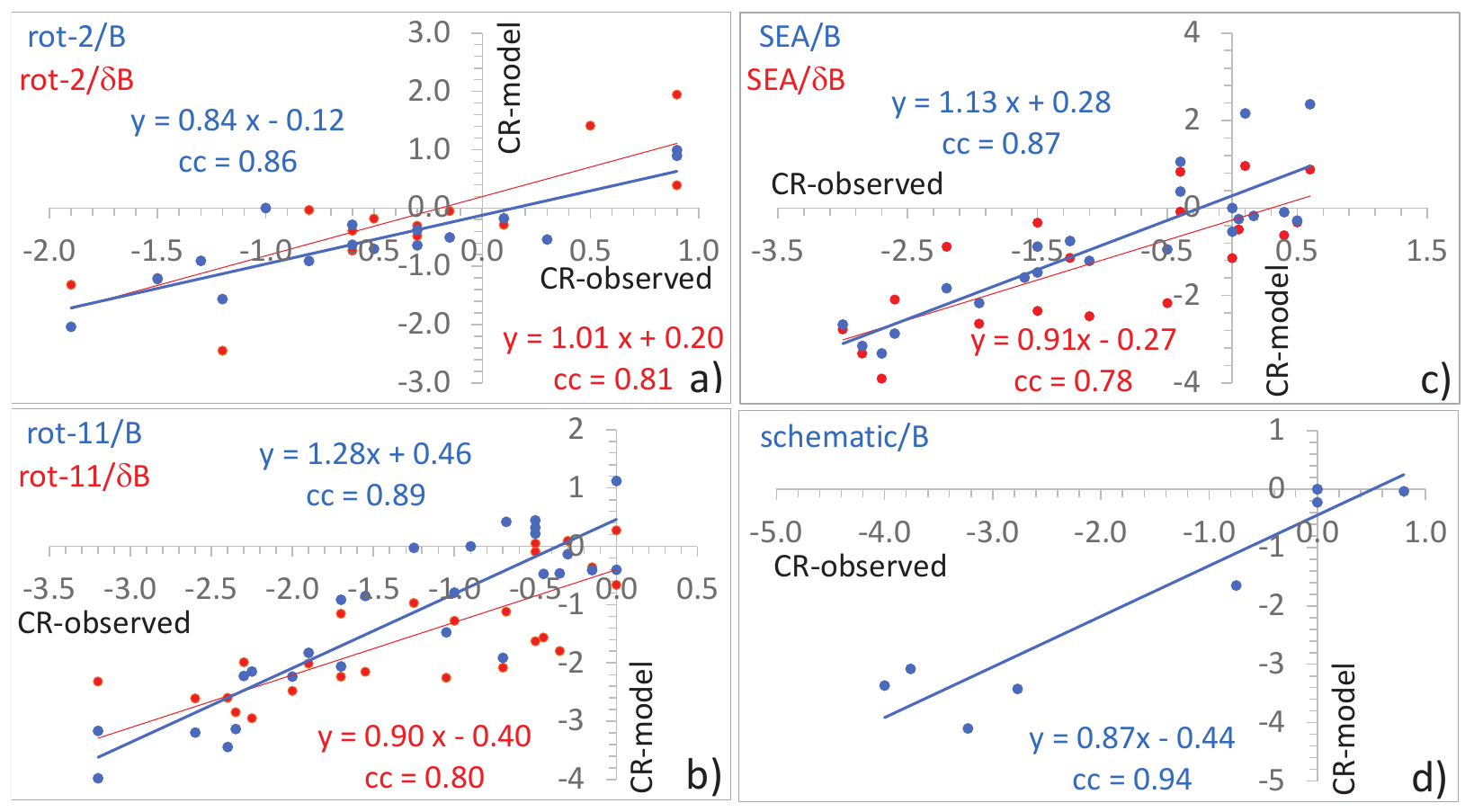}
\end{center}
\caption{
Quantitative comparison of the model values ($y$-axes) with the observed values ($x$-axes). In Panels a), b) and c) the data drawn in blue and red represent the model results based on the normalized magnetic field strength $b$ and its normalized fluctuations $\delta b$, respectively. In Panel d)  the outcome for generic schematic profile based on the $b$ model option is displayed. In the insets, linear least-squares fits are written, together with the corresponding correlation coefficients, $cc$.
}
\label{f6}
\end{figure*}

In Fig.~\ref{f6} the model results ($y$-axes) are plotted versus the observed results ($x$-axes) for the rot-2, rot-11, SEA, and generic cases.
The slopes of the linear least-squares fits, $y=c_1x+c_0$, are found to be $c_1\approx 1$ on average, and the corresponding $y$-axis intercept is $c_0\approx 0$ (a perfect match corresponds to $c_1=1$, $c_0=0$).
The correlation coefficients are high; they range from $cc=0.78$ to 0.94. The results displayed in Fig.~\ref{f6} show that the correlation coefficients are better for the $b$-option than for the $\delta b$-option, which is expected because of the various effects that have to be taken into account (high noise in the $\delta b$-data, a strong dependence on the time resolution, the unknown spatial scales and timescales of the fluctuations that affect $CR$, etc.). However, we note that the clear correlation between the model results and the observations for the $\delta b$-option is consistent with the hypothesis that the magnetic filed fluctuations govern the GCR diffusion term of the GCR transport equation, as expected from the theoretical point of view \citep[][]{parker65}. We note that $\delta B\propto B$ (see Section~\ref{discus}), which basically indicates that the fluctuations are of comparable normalized values, meaning that the turbulence attains larger amplitudes in stronger fields.
   The relative values remain comparable, however, $\delta B/B\approx 0.1$.

In Fig.~\ref{f5}e the $CR$ over-recovery is evident in both the observed and modeled values for rot-2, which is also reflected in Fig.~\ref{f6}a (dots in the top right quadrant; $x>0$, $y>0$, respectively). Moreover, the model reproduces the secondary dip of $CR$ in the case of rot-11 (Fig.~\ref{f5}f) very well.

\begin{figure*}
\begin{center}
\includegraphics[scale=0.95]{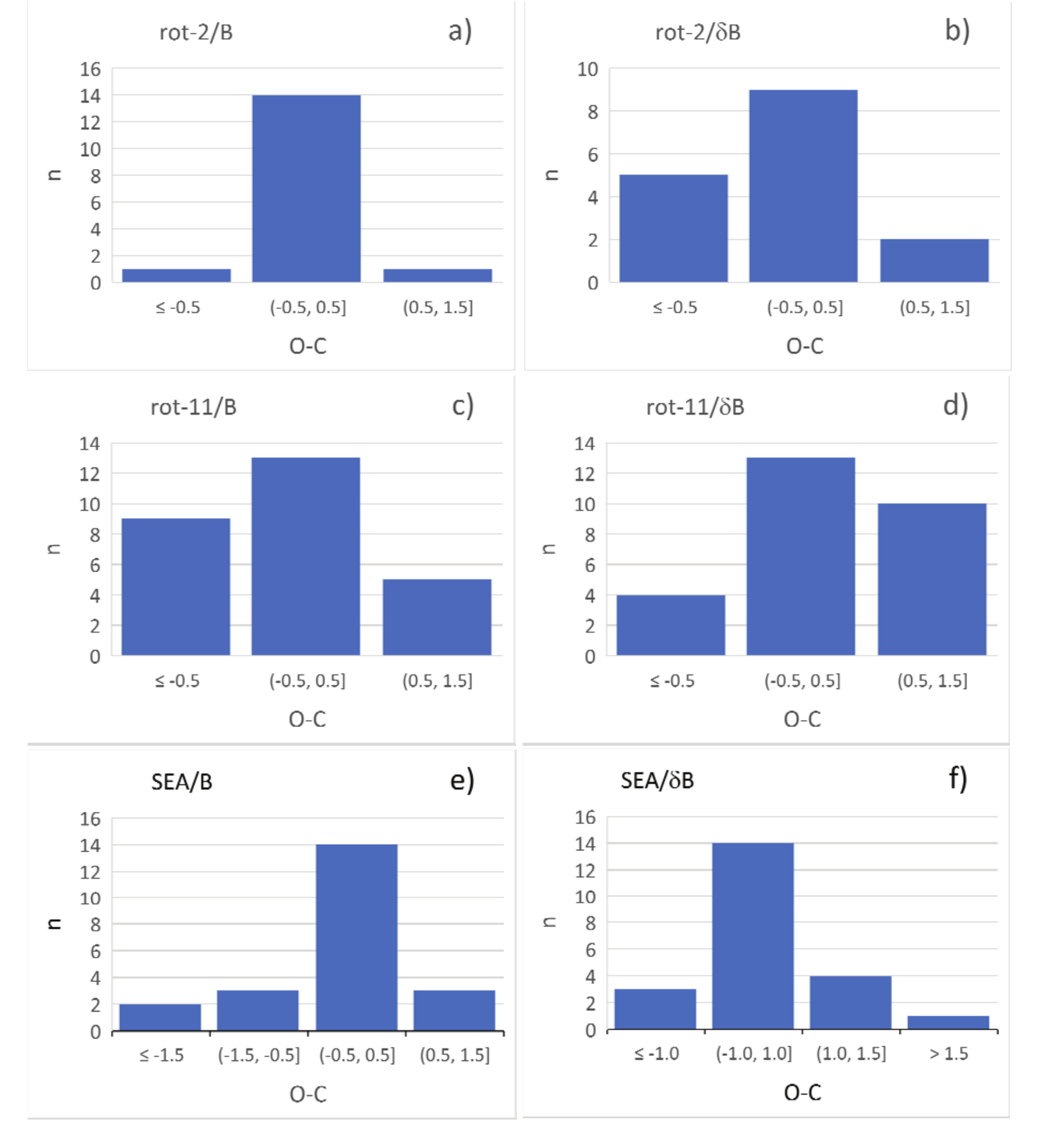}
\end{center}
\caption{
Distributions of observed-minus-calculated ($O-C$) values: Panel a): rot-2 modeled using the normalized magnetic field values $b$. Panel b): rot-2 modeled using the normalized magnetic field fluctuations $\delta b$. Panels c) and d): the same as in a) and b), but for rot-11. Panels e) and f): the same as in Panels a) and b), but now for the SEA results.
}
\label{f7}
\end{figure*}

In Fig.~\ref{f7} we present the distributions of the observed-minus-calculated $CR$ values, $O-C$, for rot-2, rot-11, and SEA for model results based on the $b$- and $\delta b$-option. In all cases, the $O-C$
values of the $CR$ instantaneous amplitudes (expressed in percent; see Fig.~\ref{f5})  peak in the range from --0.5 to +0.5\,\%. The best results are obtained for the rot-2/b-option and SEA/b-option (Figs.~\ref{f7}b and e, respectively).

\begin{table*}
\caption{ Analysis of the results
}
\label{tab}
\begin{center}
\begin{tabular}{cccccc}
  \hline
 method  &  $O-C$ (\%)  average  &  $O-C$(\%) median &  $c_1$ & $c_0$ & $cc$ \\
  \hline
rot-2/$b$, $k=0.65$ & 0.04$\pm$0.41 & 0.09 & 0.84$\pm$0.13  & -0.12$\pm$0.12 & 0.86 \\
rot-2/$\delta b$, $k=0.30$ &  -0.19$\pm$0.61 & -0.24 & 1.01$\pm$0.20  &  0.20$\pm$0.19  & 0.81 \\
rot-11/$b$, $k=0.90$ & -0.10$\pm$0.69  &  -0.08 & 1.28$\pm$0.13  &  0.46$\pm$0.21 & 0.89 \\
rot-11/$\delta b$, $k=0.75$ &  0.27$\pm$0.68  & 0.28 &  0.90$\pm$0.14   &  -0.40$\pm$0.22 & 0.80 \\
SEA/$b$, $k=0.80$ &  -0.16$\pm$0.79 & 0.05  & 1.13$\pm$0.14   &  0.28$\pm$0.22   & 0.87 \\
SEA/$\delta b$, $k=0.85$ & 0.19$\pm$0.90 & 0.17  & 0.91$\pm$0.17   &  -0.28$\pm$0.26   & 0.78 \\
schematic/b, $k=1.5$ &  0.27$\pm$0.70 & 0.44 &  0.87$\pm$0.11  &  -0.44$\pm$0.27 & 0.94  \\
  \hline
mean&   0.05$\pm$0.20   &  0.10$\pm$0.22  & 0.99$\pm$0.16  & -0.04$\pm$0.36 &    0.85$\pm$0.06     \\
  \hline
\end{tabular}
\end{center}
Mean $O-C$ values with the corresponding standard deviations and median $O-C$ values, the slopes $c_1$, the $y$-axis intercepts $c_0$ , and the correlation coefficients of the  linear least-squares fits $y=c_1x+c_0$ , where $x$ represents the observed and $y$ the modeled values. The last row shows the mean value and the corresponding standard deviation for each column.
\end{table*}

In Table~\ref{tab} we summarize the basic results of our comparison of the model results and observations. In Column 1 all the considered options are specified. In Columns 2 and 3 the average and median of the observed-minus-calculated values are shown. In Columns 4\,--\,6 the basic parameters of the linear least-squares fit are presented (the slope, $c_1$, the $y$-axis intercept, $c_0$, and the correlation coefficient, $cc$, respectively). In the bottom row, the mean values for each column are displayed. The average and median values show a good overall match between the calculated and the observed values (the mean values presented in the last row are 0.05\,$\pm$\,0.20\,\% and 0.10\,$\pm$\,0.22\,\%, respectively). Generally, the slopes of the calculated-versus-observed linear least-squares correlations are grouped around $c_1=1$, with the mean value 0.99\,$\pm$\,0.16. The $y$-axis intercept, $c_0$, ranges from --0.44 to +0.46, with the mean value of --0.04\,$\pm$\,0.36, that is, $c_0\approx 0$. The correlation coefficients range from $cc=0.78$ to 0.94, with a mean value of 0.85\,$\pm$\,0.06.

\section{Discussion}
\label{discus}

First of all,  we emphasize that our results for the overall generic data, as well as for the two specific examples that were analyzed in more detail, are related to only one long-lasting CIR. Thus, the results we obtained should not be {\it \textup{a priori}} generalized to all CIR events. This can be seen already from the fact that the scaling factor $k$ is quite different in the case of rot-2 and rot-11, as well as in comparison with the SEA and schematic generic profiles. Still, the value of $k$ for the $b$-options  is  approximately the value of $k\approx$\,1 on average (from the four inferred values of $k$, see Table~\ref{tab}, the mean value is $\overline k_b=0.96\pm0.37$). The values of $k$ for the $\delta b$-option seem to be somewhat lower ($k_{\delta b}=0.30$, 0.75, and 0.85 for rot-2, rot-11, and SEA, respectively), with a mean value of $\overline k=0.63\pm0.29$, but they are still within the same order of magnitude.

\begin{figure}
\begin{center}
\includegraphics*[scale=0.50]{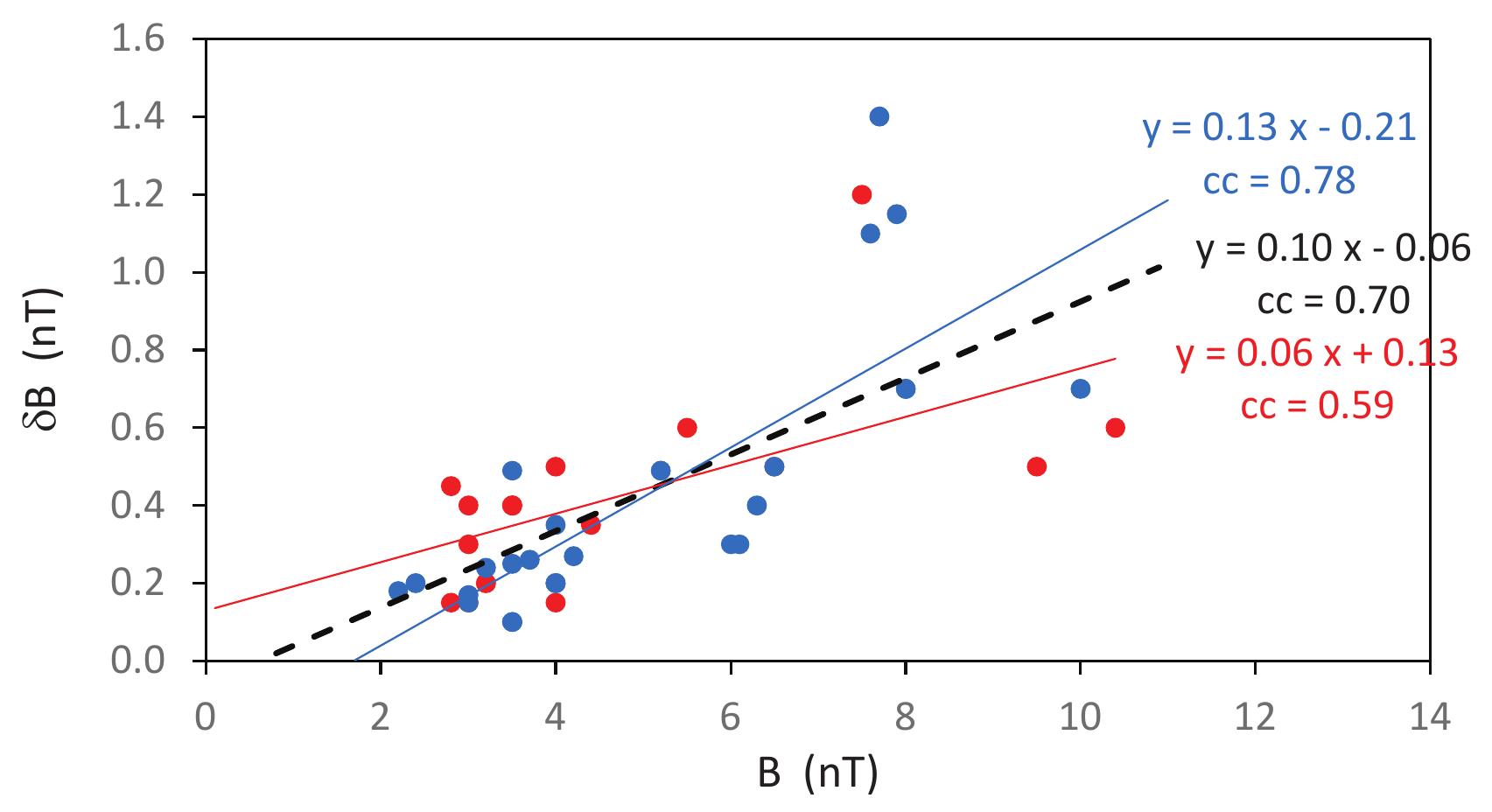}
\end{center}
\caption{
Dependence of the level of the magnetic field fluctuations, $\delta B$, on the related magnetic field strength, $B$, for rot-2 (red) and rot-11 (blue), together with the corresponding linear least-squares fit. The fit embracing all data points is shown by the dashed black line.
}
\label{f8}
\end{figure}

The dependence $\delta B(B)$ for rot-2 and rot-11, presented in Fig.~\ref{f8}, shows that $\delta B$ and $B$ are correlated (the correlation coefficients are $cc=0.59$ and 0.78, respectively).
    Although the number of data points is relatively small ($n=$\,15 and 27, respectively), the F-test confidence is high ($P=$\,98.5\% and $P\gg$\,99\%, respectively).
When the data for both data samples are put together, the correlation coefficient adds up to $cc=0.70$, with an F-test confidence $P$\,$\gg$\,99\,\%.

The linear least-squares fit for rot-2 results in $\delta B =(0.06\pm0.02)\,B + (0.13\pm0.12)$, for rot-11 in $\delta B =(0.13\pm0.02)\,B - (0.21\pm0.11)$, and for the joined data, the relation becomes $\delta B =(0.098\pm0.015)\,B - (0.06\pm0.08)$. The slopes of $c_1\approx0.1$ and the $y$-axis intercepts of $c_0\approx0$ confirm the close correlation of these two magnetic field parameters, where the field fluctuations $\delta B$ achieve about 10\,\% of the instantaneous magnetic field strength $B$ on average. The tight correlation of $\delta B$ and $B$ explains why  the model $b$-option and $\delta b$-option both work well and why the scaling factor $k$ in both options has a quite similar value.

At this point, it is appropriate to discuss the effect of the magnetic field fluctuations $\delta B$ on the GCR particle propagation. The OMNI-reported fluctuations $\delta B$ represent the
root mean square of the average magnetic field magnitude within the time interval defined by the time resolution\footnote{https://omniweb.gsfc.nasa.gov/html/ow$\_$data.html$\#$3}, in this case,  1\,h. The measurements show that during the increased-$B$ phase, the level of the fluctuations most often varyies from 0.2 to 2\,nT, for example. The typical duration of these episodes of increased fluctuations is $\delta t=1-5$\,h. The length-scale equivalent, $\delta x$, can be calculated as $\delta x=v_{\rm rot}\delta t$, where $v_{\rm rot}$ represents the velocity of the CIR rotation at 1\,au. In this way, we obtain $\delta x\approx0.01-0.05$\,au, that is, $\delta x\approx1.5-7\times10^6$\,km, and this is somewhat larger than a typical 1\,GeV proton gyroradius in the 5\,--\,10\,nT environment. This means that these fluctuation clumps taken as an entity should not affect the propagation of these particles too much \citep[][]{parker65}, and thus should not affect the diffusion coefficient significantly. The considered fluctuation clumps are a result of many fluctuations on still smaller spatial scales that would also affect the propagation of 1\,GeV protons, that is, the related diffusion coefficient and thus the GCR count rate, and this would need an additional, more careful comparison of the $CR$ behavior with $\delta B$ on different timescales.

Finally, we comment on the pre-CIR density increase and the associated $V$ depletion (Figs.~\ref{f1}c and a, respectively; see also Fig.~\ref{f2}b). It can be speculated that a possible explanation might be related to the plasma pressure gradient within the frontal part of the CIR. This pressure gradient should cause a component of the flow speed along the field lines out of the frontal boundary of CIR that is compressed and thus carries additional plasma  ahead of the CIR, that is, causes a weak pre-CIR density increase. As this flow component has a radial component that opposes the overall pre-CIR solar wind flow, this effect should also lead to the formation of the depletion in the radial speed time profile. In the situation shown in Fig.~\ref{f1}e, where there is no significant pre-CIR $V$ depletion, there is no pre-CIR density increase either.

This scenario might be also related to the formation of the pre-CIR nose in the $CR$ time profile. Namely, Fig.~\ref{f2}b shows that the nose occurs during the period of $V$ depletion (see also Fig.~\ref{f1}d). According to the previously proposed model, a lower $V$, that is, a reduced convective effect, should lead to an increase in $CR$. The situation presented in Figs.~\ref{f1}e--h is somewhat different because the nose is still observed in $CR$, although there is no pre-CIR $V$ depletion. On the other hand, inspecting the $B$ and $\delta B$ graphs in Fig.~\ref{f1}f, we find that in the period of the nose-related $CR$ increase, there is a depletion in both $B$ and $\delta B$, and according to the model, this should enable a more efficient diffusion and thus the $CR$ increase.

    Because the proposed explanation for this weak increase in GCR flux before the main FD decrease is quite speculative and cannot be fully confirmed by our analysis, we should also consider another possible explanation that is related to the mirror effect at a steep gradient of the magnetic field (including the situation in which a forward shock has already formed)  in the region in which the fast solar wind compresses the plasma in the interaction with slow solar wind. In this respect, we note that in the situation in which cylindrical geometry cannot be applied and the adiabatic approximation is not valid, the mirror effect is quite complex and cannot easily be included in the hydrodynamical model such as employed here, that is, the numerical approach has to be applied (see Kirin et al 2020). The results presented by \citet{kirin20} show that a certain population of particles tends to stay bound to the shock for a certain period of time, indicating that this effect may explain the weak increase in GCR flux before the main FD decrease. Related to the compression in the frontal parts of CIRs, we also emphasize the so-called snowplow effect \citep[e.g.,][]{guo21}, which is observed as the plasma density increase that starts even before the solar wind speed starts to rise, and which may also cause the GCR pre-CIR increase \citep[][]{guo21}.

   Finally, we note that although our model is based on 2D approach, it uses a realistic CIR structure in the equatorial plane as a structural background for a quite simple analytical model for the GCR modulation, which is based on the simplified Parker transport equation. Our model is similar to the steady-state diffusion-convection model by \citet{richardson96}, which in addition considers the adiabatic deceleration and longitudinal variations in the solar wind speed. In both models, the effect of the convection due to the solar wind outward flow is locally balanced by the inward diffusion of GCRs. \citet{richardson96} concluded that the main cause of the CIR-related GCR depression is the increase in solar wind speed, that is, the enhanced outward-convection effects, whereas the enhanced magnetic field in the the frontal parts of CIR as well as the turbulence-related drifts play a less important role. In contrast, we have found that the magnetic field  and turbulence enhancement play an important role in the GCR modulation. The results of our model are on the other hand very similar to the much more sophisticated 3D numerical model by \citet{alania11}. They considered a steady-state heliosphere in which the CIR is represented as heliolongitudinal variation of solar wind speed and applied the full Parker transport equation. They found that the variation in GCR intensity is inversely correlated with the modulation parameter, which is proportional to the product of the solar wind velocity V and the strength of the interplanetary magnetic field B. This is exactly the case in our model. Moreover, their calculations have shown that changing the HCS structure in the model does not significantly affect the results
   \citep[for a detailed study of the effect of HCSs on GCRs, see, e.g.,][]{thomas14}.

    Results similar to those obtained by  \citet{alania11} were also found by \citet{guo16} using the 3D coupled MHD-CR transport model. They also found that that GCR variations depend on the ratio of the diffusion coefficients in the fast and slow solar winds, as is the case in our model. Finally, we compared our results to the most recent dedicated 3D coupled MHD-CR transport models by \citet{kopp17}  and \citet{luo20}. Their simulations show that enhanced convection, reduced diffusion, and drifts due the magnetic field enhancement may contribute to the recurrent FD. However, the simulations cannot distinguish the relative contribution of different mechanisms. Our results indicate that convection and diffusion are enough to describe the basic properties of recurrent FDs.  On the other hand, we do not exclude the possibility that drifts are needed to explain the fine structures within the recurrent FD. We also note that in all of the models above, including our model, the transport parameters need to be tuned to successfully reproduce the observations. In the quantitative aspect, our simple analytical model therefore performs equally well as the much more sophisticated and physically more realistic models.

\section{Conclusion}
\label{concl}

Our analysis explains the physical background of RFDs caused by CIR and HSS structures in terms of the convection-diffusion approach. It is demonstrated that the enhanced convection-effect caused by the increased velocity of HSS and the reduced diffusion-effect caused by the enhanced magnetic field fluctuations within the CIR and HSS structure are sufficient to explain the observations. This approach was  verified by applying a relatively simple analytical model to the observations.

The main guiding observational fact that leads in this direction is the close similarity of the $V(t)$ curve and the inverted $CR(t)$ curve (Figs.~\ref{f4} and~\ref{f5}; for more examples, see Paper 1). However, the enhanced convection effect cannot explain the $CR(t)$ behavior without an additional ingredient. After we inspected the in situ solar wind data, we realized that this has to be the increase in magnetic field strength $B$ within the CIR structure, or a physically more appropriate option, the enhanced magnetic field fluctuations $\delta B$ that reduce the CR diffusion.

The combined effect of the increased solar wind speed and the enhanced magnetic field fluctuations practically entirely explain the overall RFD behavior in the 1\,au range, related to the CIR and HSS structure at these distances. The two effects taken together  can also explain the phenomenon of the over-recovery that is sometimes observed after the RFD events. It is likely that the over-recovery occurs (based on the two employed examples) in the events in which the solar wind speed and the magnetic field fluctuations are both lower than in the pre-event state.

\begin{acknowledgements}
This work has been supported by Croatian Science Foundation under the project 7549 "Millimeter and submillimeter observations of the solar chromosphere
with ALMA". B.V. and M.D. also acknowledge support by the Croatian Science Foundation under the project IP-2020-02-9893 (ICOHOSS). B.H. acknowledges the financial support via projects  HE 3279/15-1 funded
by the Deutsche Forschungsgemeinschaft (DFG).
\end{acknowledgements}

\bibliographystyle{aa} 
\bibliography{cir}

\newpage
\begin{appendix}
\section{Segmental linear fitting}

\begin{figure}
\includegraphics[width=8cm]{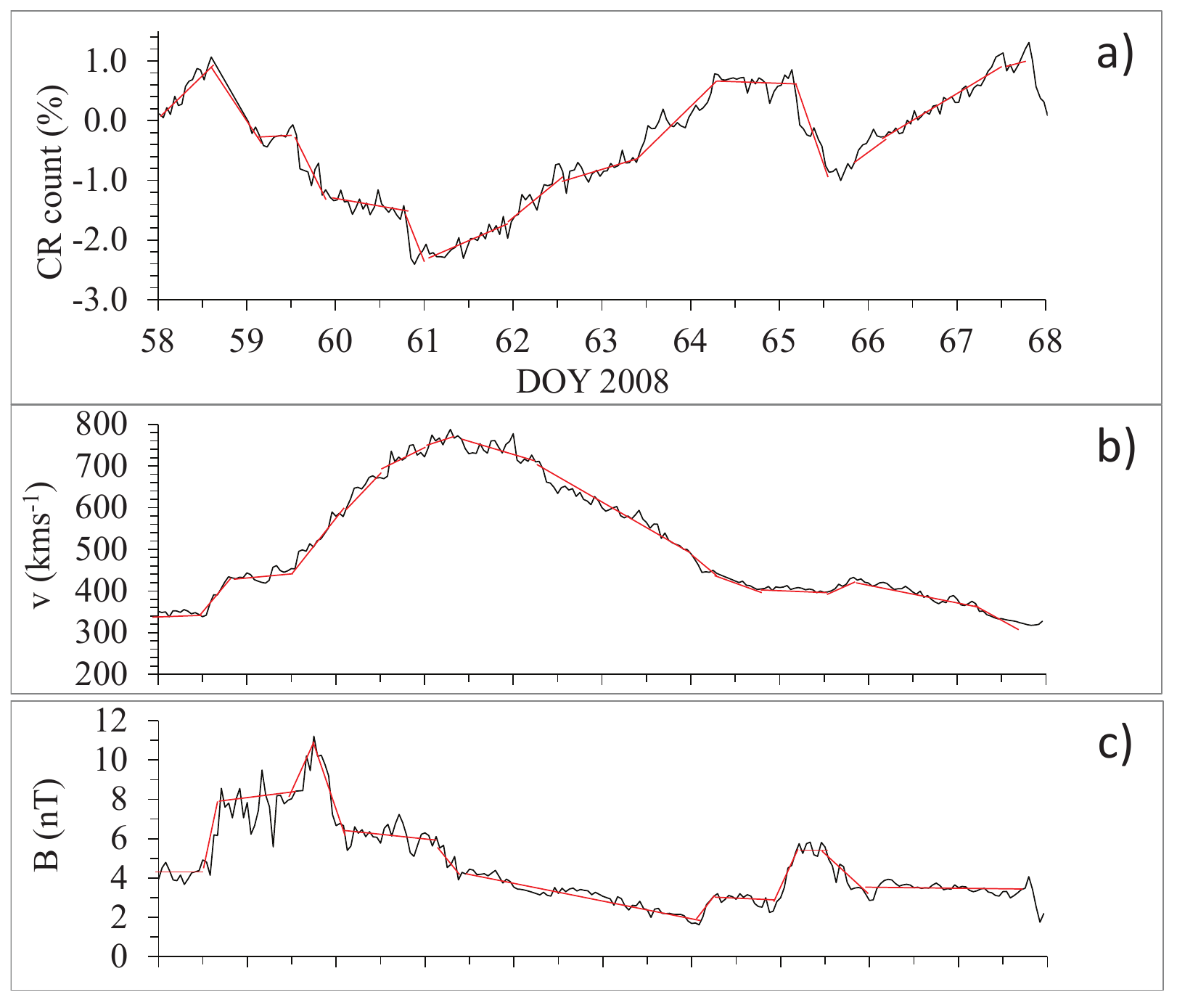}
\caption{Three examples of the SLF  (rot-11). Panel a): EPHIN GCR count rate. Panel b): solar wind velocity time profile. Panel c): interplanetary magnetic field time profile.}
\label{f9}
\end{figure}

In Fig~\ref{f9} the SLF procedure is illustrated. It is applied to the EPHIN GCR count rate, solar wind velocity time profile, and interplanetary magnetic field time profile. The procedure is as follows. First, the moments in a given curve are identified, where the slope of the curve (neglecting the noise) significantly changes (nodes). Then, the parts of the curve between nodes are fit by a linear fit (red lines).

\end{appendix}

\end{document}